\documentclass[prb,aps,twocolumn,showpacs,superscriptaddress,floatfix]{revtex4}
\usepackage{epsfig}
\usepackage{amsmath}
\usepackage{amssymb}
\usepackage{amsfonts}
\usepackage{mathptmx}
\usepackage{eucal}
\usepackage{bm}

\DeclareMathOperator{\sgn}{\mathop{\mathrm{sgn}}}
\DeclareMathOperator{\re}{\mathop{\mathrm{Re}}}

\newcommand{\Eq}[1]{Eq.~(\ref{#1})}
\newcommand{\Eqs}[1]{Eqs.~(\ref{#1})}

\begin{document}

\title{Current-voltage characteristics of tunnel Josephson junctions with a ferromagnetic interlayer}
\author{A.~S.~Vasenko}
\affiliation{Institut Laue-Langevin, 6 rue Jules Horowitz, BP 156,
38042 Grenoble, France}
\affiliation{LPMMC, Universit\'{e} Joseph
Fourier and CNRS, 25 Avenue des Martyrs, BP 166, 38042 Grenoble,
France}
\affiliation{Donostia International Physics Center (DIPC), Manuel
de Lardizabal 4, E-20018 San Sebasti\'{a}n, Spain}
\author{S.~Kawabata}
\affiliation{Nanosystem Research Institute (NRI), National
Institute of Advanced Industrial Science and Technology (AIST),
and JST-CREST, Tsukuba, Ibaraki, 305-8568, Japan}
\author{A.~A.~Golubov}
\affiliation{Faculty of Science and Technology and MESA$^+$
Institute for Nanotechnology, University of Twente, 7500 AE
Enschede, The Netherlands}
\author{M.~Yu.~Kupriyanov}
\affiliation{Nuclear Physics Institute, Moscow State University, Moscow, 119992, Russia}
\author{C.~Lacroix}
\affiliation{Institut N\'{e}el, Universit\'{e} Joseph Fourier and CNRS, 25 avenue des Martyrs,
BP 166, 38042 Grenoble, France}
\author{F.~S.~Bergeret}
\affiliation{Centro de F\'{i}sica de Materiales (CFM-MPC), Centro
Mixto CSIC-UPV/EHU, Manuel de Lardizabal 5, E-20018 San
Sebasti\'{a}n, Spain}
\affiliation{Donostia International Physics Center (DIPC), Manuel
de Lardizabal 4, E-20018 San Sebasti\'{a}n, Spain}
\author{F.~W.~J.~Hekking}
\affiliation{LPMMC, Universit\'{e} Joseph Fourier and CNRS, 25
Avenue des Martyrs, BP 166, 38042 Grenoble, France}
\date{\today}

\begin{abstract}
We present a quantitative study of  the current-voltage characteristics (CVC) of
diffusive superconductor/ insulator/ ferromagnet/ superconductor (SIFS) tunnel Josephson
junctions. In order to obtain the CVC we calculate the density of states (DOS) in the F/S
bilayer for arbitrary length of the ferromagnetic layer, using quasiclassical theory. For
a ferromagnetic layer thickness larger than the characteristic penetration depth of the
superconducting condensate into the F layer, we find an analytical expression which
agrees with the DOS obtained from a self-consistent numerical method. We discuss general
properties of the DOS and its dependence on the parameters of the ferromagnetic layer. In
particular we focus our analysis  on  the DOS oscillations at the Fermi energy. Using
the numerically obtained DOS we calculate the corresponding CVC and discuss their properties.
Finally, we use CVC to calculate the macroscopic quantum tunneling
(MQT) escape rate for the current biased SIFS junctions by taking into account the
dissipative correction due to the quasiparticle tunneling. We show that the influence of
the quasiparticle dissipation on the macroscopic quantum dynamics of SIFS junctions is
small, which is an advantage of SIFS junctions for superconducting qubits applications.
\end{abstract}

\pacs{74.45.+c, 74.50.+r, 74.78.Fk, 75.30.Et}
\maketitle


\section{Introduction}

The possibility to switch the ground-state of a Josephson junction from a $0$ to a $\pi$
phase state and the  possible application of such junctions in quantum information  led
to a renewal interest in the study of the so called $\pi$ Josephson junctions. The
existence of such a transition was predicted more than thirty years ago,\cite{Bulaevskii}
however due to technological requirements only recently it was observed. The realization
of $\pi$ Josephson junctions was achieved  in superconductor/ ferromagnet/ superconductor
(SFS) junctions. \cite{Ryazanov,
Kontos_exp, Blum, Guichard, Sellier, Bauer, Bell, Born, Shelukhin, Pepe, Oboznov, Weides, Weides2,
Weides3, Pfeiffer, Bannykh, Kemmler, Anwar, Khaire, Robinson} Microscopically, S/F
hybrid structures are characterized by an unusual proximity effect, with a  damped
oscillatory behavior of the superconducting correlations in the F layer (for a review see
Refs.~\onlinecite{RevB, RevG, RevV} and references therein). This unusual proximity
effect in S/F layered structures leads to a number of striking phenomena like the
nonmonotonic dependence of their critical temperature and the appearance of oscillations
of critical current in SFS Josephson junctions as a function of the F layer
thickness.\cite{Ryazanov, Kontos_exp} In particular the  change of sign of the critical current
corresponds to the so-called $0-\pi$ transition.

On the other hand, SFS junctions,  as any  metallic junction exhibit very small
resistances and therefore are  not quite suitable for those applications, for which
active Josephson junctions are required. This problem can be solved by adding an
additional insulating (I) layer to increase the resistance. SIFS junctions represent an
interesting case for practical use of $\pi$ Josephson junctions. For instance, a
SIFS structure offers the freedom to tune the critical current density over a wide range
and at the same time realize high values of the product of the junction critical current
$I_c$ and its normal state resistance $R_n$. \cite{Weides, Weides2, Weides3}  In
addition, Nb based tunnel junctions are usually underdamped, which is desired for many
applications. Due to these advantages,  SIFS $\pi$ junctions have been proposed
as potential elements in superconducting classical and quantum logic
circuits.\cite{logic, rf:Feofanov} For instance, SIFS junctions can be used as complementary
elements ($\pi$ - shifters) in RSFQ circuits (see Ref.~\onlinecite{RSFQ} and references
therein). Finally, SIFS structures have been proposed for
the realization of so called $\varphi$-junctions with a $\varphi$ drop in the ground
state, where $0 < \varphi < \pi$.\cite{phi} The properties of SIFS junctions have been
intensively studied both experimentally\cite{Kontos_exp,
Guichard, Born, Pepe, Weides, Weides2, Weides3, Pfeiffer, Bannykh,
Kemmler} and theoretically. \cite{phi, Vasenko, Pugach,
Volkov_SIFS} However, properties of the quasiparticle current have received
relatively little attention so far, although they can be very important for the
description of SIFS junctions as possible elements of superconducting logic circuits.

The purpose of this work is to provide a quantitative model describing the behavior of
quasiparticle current in SIFS junctions as a function of parameters characterizing
material properties of the ferromagnetic interlayer. We also focus our study on  the
properties of the density of states (DOS) in S/F bilayers and discuss the oscillations of
the DOS at Fermi energy. Finally, we calculate the macroscopic quantum tunneling (MQT)
escape rate for current-biased SIFS junctions by taking into account the dissipative
correction due to the quasiparticle tunneling. Based on this we conclude that the
influence of the quasiparticle dissipation on the macroscopic quantum dynamics of SIFS
junctions is small, which is an advantage of SIFS junctions for quantum logic
(qubit) applications.

The paper is organized as follows. In the next section we formulate the theoretical model
and the basic equations. In Sec.~\ref{DOS_FS} we solve the  nonlinear Usadel equations
numerically for arbitrary length of the ferromagnetic layer, and calculate the DOS in the
F layer. We compare these results with an analytical expression  for DOS in case of  a
long SIFS junction, {\it i.e.} when the thickness  $d_f$ of the ferromagnetic layer is
much larger than the decay length of the characteristic superconducting
correlations in the ferromagnet $ \xi_{f1}$.   We  also discuss the oscillations of the
DOS at the Fermi energy. In Sec.~\ref{CVC_SIFS} we present the current-voltage
characteristics of SIFS junctions for different parameters of the ferromagnetic
interlayer. In Sec.~\ref{sec-MQT}, we use these data to calculate the MQT escape rate for
current-biased SIFS junctions by taking into account the dissipative effect of the
quasiparticle tunneling. Finally  we summarize the results in Sec.~\ref{Concl}.


\section{Model and basic equations}\label{Model}

\begin{figure}[tb]
\epsfxsize=8.5cm\epsffile{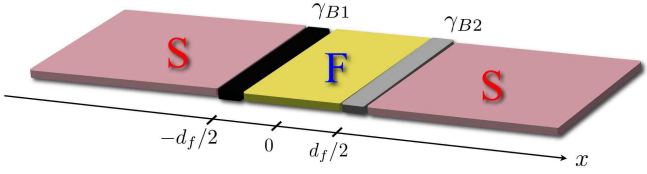} \vspace{-3mm}
\caption{(Color online) Geometry of the considered system. The thickness of the
ferromagnetic interlayer is $d_f$. The transparency of the left  (right) S/F interface is
characterized by the coefficient $\protect\gamma_{B1(B2)}$.  The left interface is an
insulating barrier, $\protect\gamma_{B1} \gg 1$ (shown by a black line), while the right
interface is transparent, $\protect\gamma_{B2} \ll 1$ (shown by a grey line).}
\label{SIFS}\vspace{-5mm}
\end{figure}

We consider a SIFS junction such as the one depicted in Fig.~\ref{SIFS}. It consists of a
ferromagnetic layer of thickness $d_{f}$ and two thick superconducting electrodes along
the $x$ direction. The left and right superconductor/ ferromagnet interfaces are
characterized by the dimensionless parameters $\gamma _{B1}$ and $\gamma _{B2}$,\cite{KL, Chalmers}
respectively, where $\gamma _{B1,B2}=R_{B1,B2}\sigma _{n}/\xi _{n}$, $R_{B1,B2}$ are the
resistances of the left and right S/F interfaces, respectively, $\sigma _{n}$ is the
conductivity of the F layer, $\xi _{n}=\sqrt{D_{f}/2\pi T_{c}}$, $D_{f}$ is the diffusion
coefficient in the ferromagnetic metal and $T_{c}$ is the critical temperature of the
superconductor (we assume $\hbar =k_{B}=1$, except for  Sec.~\ref{sec-MQT}). We also
assume that the S/F interfaces are not magnetically active. We will consider the
diffusive limit, in which the elastic scattering length $\ell $ is much smaller than the
decay characteristic length $\xi _{f1} = \min\left[ \xi_{f1 \uparrow}, \xi_{f1
\downarrow} \right]$ (for the definitions of $\xi_{f1 \uparrow (\downarrow)}$ see
\Eqs{xif12} below).

We assume that the  tunneling barrier is located at the left
S/F interface, while the right interface is perfectly transparent,
this means that  $\gamma_{B1} \gg 1$  while $\gamma_{B2} \ll 1$.
In this case the left S layer and the right F/S bilayer in
Fig.~\ref{SIFS} are decoupled and we can calculate the
quasiparticle current through a SIFS junction using the standard
 tunneling formula\cite{Werthammer}
\begin{equation}\label{I(V)}
I = \frac{1}{e R} \int_{-\infty}^{\infty}dE N_s(E - eV) N_f(E)
\left[ f(E - eV) - f(E) \right],
\end{equation}
where $N_s(E) = |E|\Theta(|E|-\Delta)/\sqrt{E^2 - \Delta^2}$ is
the BCS DOS, $\Theta(x)$ is the Heaviside step function, $N_f(E)$
is the DOS in the ferromagnetic interlayer at $x=-d_f/2$,
$f(E)=[1+\exp(E/T)]^{-1}$ is the Fermi function, and $R \equiv
R_{B 1}$. Both $N_s(E)$ and $N_f(E)$ are normalized to their
values in the normal state. In particular at zero temperature, $T=0$, the
current  acquires the form,
\begin{equation}\label{I(V)T0}
I =\Theta(eV-\Delta) \frac{1}{e R} \int_{0}^{eV-\Delta}dE N_s(E -
eV) N_f(E)\; .
\end{equation}
To obtain $N_f(E)$ we notice that since $\gamma_{B1} \gg 1$, the
left superconducting lead does not influence the DOS in the
ferromagnetic interlayer (to zero order in the barrier
transparency). This  reduces the  problem to the following: we
need to find the DOS of a single F/S bilayer, which can be done by
solving the Usadel equations in the ferromagnetic layer.

Using the $\theta$-parameterizations of the normal and
anomalous Green functions, $G=\cos \theta $, $F=\sin \theta$, we
can write the Usadel equations in the F layer as \cite{Usadel,
Demler, Gusakova}
\begin{align}\label{Usadel_magn}
\frac{D_{f}}{2} \frac{\partial ^{2}\theta _{f\uparrow (\downarrow
)}}{\partial x^{2}} = &\left( \omega \pm ih + \frac{1}{\tau
_{z}}\cos \theta_{f\uparrow (\downarrow )}\right) \sin
\theta_{f\uparrow (\downarrow )}\nonumber
\\
+ &\frac{1}{\tau_{x}} \sin\left( \theta_{f\uparrow} +
\theta_{f\downarrow} \right) \pm \frac{1}{\tau_{so}} \sin\left(
\theta_{f\uparrow} - \theta_{f\downarrow} \right),
\end{align}
where the positive  and negative signs correspond to the spin up
$\uparrow $ and spin down  $\downarrow $ states respectively. In
this notation the spin up state corresponds to the anomalous Green
function $F_\uparrow \sim \langle \psi_\uparrow \psi_\downarrow
\rangle$ while the spin down state corresponds  to $F_\downarrow
\sim \langle \psi_\downarrow \psi_\uparrow \rangle$, where
$\psi_{\uparrow (\downarrow)}$ are the electron fermionic
operators. The $\omega =2 \pi T(n +\frac{1}{2})$ are the Matsubara
frequencies, and $h$ is the exchange field in the ferromagnet. The
scattering times are labelled here as $\tau_z$, $\tau_x$ and
$\tau_{so}$, where $\tau_{z(x)}$ corresponds to the magnetic
scattering parallel (perpendicular) to the quantization axis, and
$\tau_{so}$ is the spin-orbit scattering time.\cite{Faure,
Bergeret, IvanovSF, AG}

We consider here  ferromagnets with a strong uniaxial anisotropy,
in which case the magnetic scattering does not couple the spin up
and spin down electron populations, {\it i.e.} the perpendicular
fluctuations of the exchange field are suppressed ($\tau_x^{-1}
\sim 0$). Therefore, we will neglect $\tau_x$ in our consideration
and denote $\tau_z$ as a magnetic scattering time $\tau_m$. We
will also consider ferromagnets with weak spin-orbit interactions
and henceforth also neglect the spin-orbit scattering time
$\tau_{so}$.  In this case  the Usadel equations in the
ferromagnetic layer for different spin projections are not coupled
any more and can be written as,
\begin{equation}\label{Usadel}
\frac{D_{f}}{2} \frac{\partial ^{2}\theta _{f\uparrow (\downarrow
)}}{\partial x^{2}} = \left( \omega \pm ih + \frac{\cos
\theta_{f\uparrow (\downarrow )}}{\tau _{m}}\right) \sin
\theta_{f\uparrow (\downarrow )},
\end{equation}
while in the S layer the Usadel equations take the form
\begin{equation}\label{Usadel_S}
\frac{D_s}{2} \frac{\partial^2 \theta_s}{\partial x^2} = \omega
\sin \theta_s - \Delta(x) \cos \theta_s.
\end{equation}
Here $D_s$ is the diffusion coefficient in the superconductor and
$\Delta(x)$ is the superconducting pair potential. Notice that in
the latter equation  we have omitted the subscripts `$\uparrow
(\downarrow)$' because both equations are identical in the
superconductor.

\Eqs{Usadel}, \eqref{Usadel_S} should be complemented by the
self-consistency equation for the superconducting order parameter
$\Delta$,
\begin{equation}
\Delta (x)\ln \frac{T_c}{T} = \pi T \sum\limits_{\omega > 0} \left( \frac{2\Delta
(x)}{\omega}-\sin \theta _{s \uparrow} - \sin \theta _{s \downarrow} \right)
\label{Delta},
\end{equation}
and by the boundary conditions at the  outer  boundary of the
ferromagnet
\begin{equation}\label{leftBK}
\left(\frac{\partial \theta_f}{\partial x} \right)_{-d_f/2} = 0,
\end{equation}
and at the  F/S interface,\cite{KL}
\begin{subequations}
\label{KL}
\begin{align}
\xi_n\gamma\left( \frac{\partial \theta_f}{\partial x} \right)_{d_f/2} &=
\xi_s \left( \frac{\partial \theta_s}{\partial x} \right)_{d_f/2},
\label{KL1} \\
\xi_n \gamma_{B2} \left( \frac{\partial \theta_f}{\partial x}
\right)_{d_f/2} &= \sin\left( \theta_s - \theta_f \right)_{d_f/2},
\label{KL_DOS}
\end{align}
\end{subequations}
where $\gamma = \xi_s\sigma_n/\xi_n\sigma_s$, $\sigma_s$ is the conductivity of the S
layer and $\xi_s = \sqrt{D_s/2\pi T_c}$. The parameter $\gamma$ determines the strength
of suppression of superconductivity in the right S lead near the interface compared to
the bulk: no suppression occurs for $\gamma=0$, while strong suppression takes place for
$\gamma \gg 1$. In our numerical calculations we will assume small $\gamma \ll 1$. Notice
that the interface parameters do not depend on the spin direction. In other words we are
not considering spin-active interfaces.  In the case of spin-active barriers, one
should use the boundary conditions introduced in Refs.~\onlinecite{CottetBelzig, Cottet,
CottetLinder}, rather than the standard Kupriyanov-Lukichev boundary conditions
[\Eqs{KL}].

To complete the boundary problem we also set a boundary condition
at $x = \infty$,
\begin{equation}\label{gran1}
\theta_s(\infty) = \arctan\frac{\Delta}{\omega},
\end{equation}
where the Green functions acquire the well known bulk BCS form. \Eqs{Usadel}-\eqref{gran1}
represent a closed set of equations that should be solved
self-consistently.  As it will be discussed in the next section, the knowdlege of the
Green function will allow us to compute the DOS at the outer F boundary.


\section{Density of states in the F/S bilayer}\label{DOS_FS}

The DOS $N_f(E)$ normalized to the DOS in the normal state, can be
written as
\begin{equation}
N_f(E) = \left[ N_{f \uparrow}(E) + N_{f \downarrow}(E)\right]/2,
\label{DOS_full}
\end{equation}
where $N_{f \uparrow(\downarrow)}(E)$ are the spin resolved
DOS written in terms of spectral angle $\theta$,
\begin{equation}
N_{f \uparrow(\downarrow)}(E) =
\re\left[\cos\theta_{f \uparrow(\downarrow)}(%
i\omega \rightarrow E + i0)\right].\label{DOS_spin}
\end{equation}
To obtain $N_f$, we use a self-consistent two-step iterative
procedure\cite{Golubov1,GK1,GK2,Gusakova}. In the first step we calculate the pair
potential coordinate dependence $\Delta(x)$ using the self-consistency equation in the S
layer, Eq.~(\ref{Delta}). Then, by proceeding to the analytical continuation in
Eqs.~(\ref{Usadel}),~\eqref{Usadel_S} over the quasiparticle energy $i\omega \rightarrow
E + i0$ and using the $\Delta(x)$ dependence obtained in the previous step, we find the
Green functions by repeating the iterations until convergency is reached.

Before showing the numerical results we consider an analytic limiting case. If the F
layer is thick enough ($d_f \gg \xi_{f1}$) and  $\gamma = 0$ in \Eq{KL1}, the DOS at the
free boundary of the ferromagnet can be written as \cite{Cretinon, Vasenko}
\begin{equation}  \label{DOS_bound}
N_{f \uparrow(\downarrow)}(E) = \re[ \cos\theta_{b
\uparrow(\downarrow)} ] \approx 1 - \frac{1}{2}\re \theta_{b
\uparrow(\downarrow)}^2.
\end{equation}
Here $\theta_{b \uparrow(\downarrow)}$ is the  value
of $\theta_f$ at $x=-d_f/2$, given by
\begin{equation}  \label{theta_bound}
\theta_{b \uparrow(\downarrow)}= \frac{8
F(E)}{\sqrt{(1-\eta^2)F^2(E) + 1} + 1} \exp\left(
-p\frac{d_f}{\xi_f}\right),
\end{equation}
where $\xi_f = \sqrt{D_f/h}$. In \Eq{theta_bound} we use the
following notations,
\begin{subequations}
\label{qef}
\begin{align}
p_{\uparrow(\downarrow)} &= \sqrt{2/h}\sqrt{-iE_R \pm ih + 1/\tau_m}, \label{p}\\
\eta^2_{\uparrow(\downarrow)} &= (1/\tau_m)(-iE_R \pm ih + 1/\tau_m)^{-1}, \label{eta}\\
F(E) &= \frac{\Delta}{-iE_R + \sqrt{\Delta^2 - E_R^2}}, \quad E_R
= E + i0.
\end{align}
\end{subequations}
Here, we again adopt the convention that a positive (negative) sign in front of
$h$ corresponds to the spin up state $\uparrow$ (spin down state $\downarrow$).
Hereafter we will write spin labels $\uparrow(\downarrow)$  explicitly only when needed.

From \Eqs{DOS_bound}-\eqref{theta_bound} we obtain for the full
DOS the following expression in the limit  $d_f \gg \xi_{f1}$,
\begin{equation}
N_f \approx 1 - \re \sum_{\uparrow, \; \downarrow} \frac{16 F^2(E)
\exp\left( -p\frac{2d_f}{\xi_f}\right)}{(\sqrt{(1-\eta^2)F^2(E) +
1} + 1)^2}.\label{Napprox}
\end{equation}
At this point, we define the characteristic decay
and oscillation lengths  $\xi_{f 1,2 \uparrow (\downarrow)}$ as
\begin{subequations}\label{xif12}
\begin{align}
p_{\uparrow (\downarrow)}/\xi_f &= 1/\xi_{f1 \uparrow
(\downarrow)} + i \sgn(h \mp E) /\xi_{f2 \uparrow (\downarrow)},
\\
\frac{1}{\xi_{f1 \uparrow (\downarrow)}} &= \frac{1}{\xi_f}
\sqrt{\sqrt{\left( \frac{E \mp h}{h} \right)^2 + \frac{1}{h^2
\tau_m^2}} + \frac{1}{h \tau_m}},\label{x_dec}
\\
\frac{1}{\xi_{f2 \uparrow (\downarrow)}} &= \frac{1}{\xi_f}
\sqrt{\sqrt{\left( \frac{E \mp h}{h} \right)^2 + \frac{1}{h^2
\tau_m^2}} - \frac{1}{h \tau_m}}.\label{x_osc}
\end{align}
\end{subequations}
In the absence of magnetic scattering  both  lengths coincide and are equal to $\xi_f
\sqrt{h/|E \mp h|}$ for different spin orientations. One can rewrite \Eq{Napprox} in the
following form,
\begin{align}
N_f \approx 1 &- \sum_{\uparrow, \; \downarrow}\exp\left( -\frac{2
d_f}{\xi_{f 1}} \right) \biggl[\mathcal{A} \sin\left( \chi +
\frac{2 d_f}{\xi_{f 2}} \right)\nonumber
\\
&+ \mathcal{B} \cos\left( \chi + \frac{2 d_f}{\xi_{f 2}}
\right)\biggr],\label{NapproxAB}
\end{align}
where the coefficients $\mathcal{A}$, $\mathcal{B}$ and $\chi$ can be obtained
by expansion of the real part in \Eq{Napprox}; only two of them are independent. This
form explicitly shows the damped oscillatory behavior of superconducting correlations in
the F layer. The  lengths $\xi_{f 1,2}$ are also the lengths of decay and oscillations of
the critical current in SIFS junctions (see Eqs.~(26) in Ref.~\onlinecite{Vasenko}). The
period of the DOS oscillations is approximately twice smaller than the period of the
critical current oscillations and the exponential decay is approximately twice faster
than the decay of the critical current.\cite{Vasenko}

\begin{figure}[tb]
\epsfxsize=8cm\epsffile{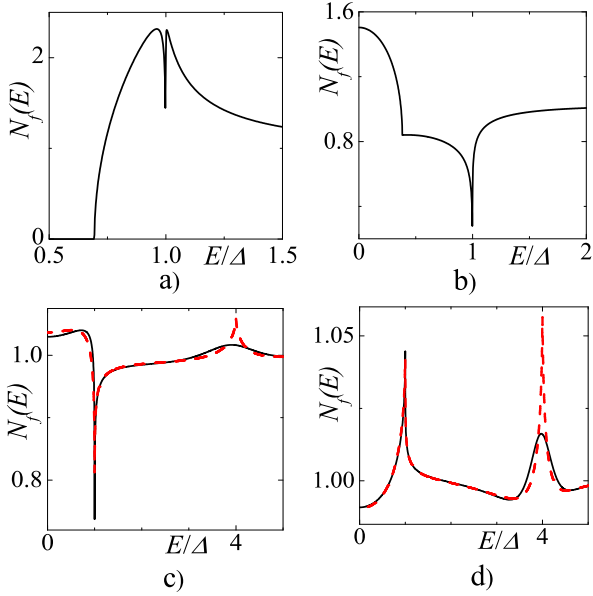} \vspace{-3mm} \caption{ (Color
online) DOS $N_f(E)$ on the free boundary of the F layer in the
F/S bilayer calculated numerically in the absence of magnetic
scattering ($1/\tau_m \Delta = 0$) for different values of the F
layer thickness $d_f$, $h/\Delta = 4$, $T = 0.1T_c$. Parameters of
the F/S interface are $\gamma = \gamma_{B2} = 0.01$.
(a): $d_f/\protect\xi_n = 0.5$, (b): $d_f/\protect\xi_n = 1$, (c): $%
d_f/\protect\xi_n = 2$, (d): $d_f/\protect\xi_n = 3$. The
approximate analytical solution, \Eq{Napprox}, is shown by dashed
red lines.} \label{DOS_a0} \vspace{-5mm}
\end{figure}

\begin{figure}[tb]
\epsfxsize=8cm\epsffile{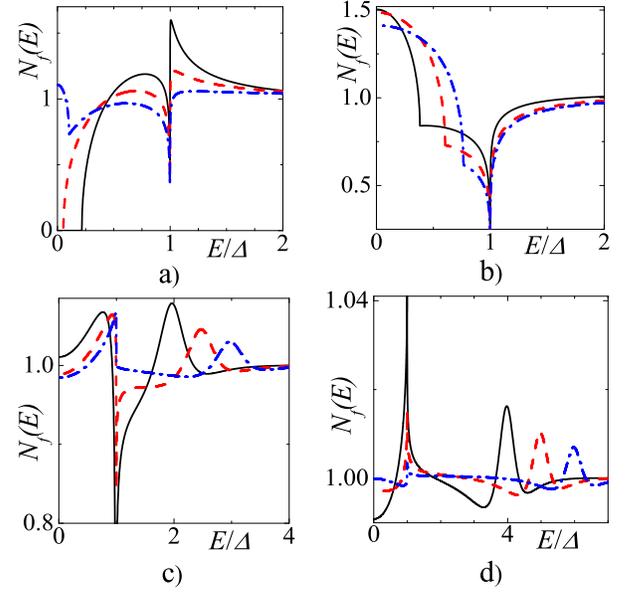} \vspace{-3mm} \caption{ (Color
online) DOS $N_f(E)$ on the free boundary of the F layer in the
F/S bilayer calculated numerically in the absence of magnetic
scattering ($1/\tau_m \Delta = 0$) for different values of the
exchange field $h$. Parameters of the F/S interface are $\gamma =
\gamma_{B2} = 0.01$, $T = 0.1T_c$. Plots (a) and (b): $d_f/\xi_n =
1$; plots (c) and (d): $d_f/\xi_n = 3$. For plots (a) and (c)
solid black line corresponds to $h/\Delta = 2$, dashed red line to
$h/\Delta = 2.5$, dash-dotted blue line to $h/\Delta = 3$. For
plots (b) and (d) solid black line corresponds to $h/\Delta = 4$,
dashed red line to $h/\Delta = 5$, dash-dotted blue line to
$h/\Delta = 6$.} \label{DOS_h} \vspace{-5mm}
\end{figure}

\begin{figure}[tb]
\epsfxsize=8cm\epsffile{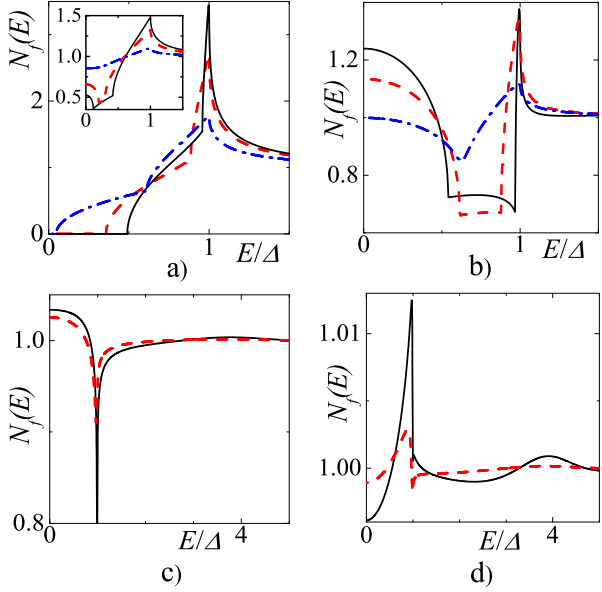} \vspace{-3mm} \caption{ (Color online) DOS $N_f(E)$ at
the free boundary of the F layer in the F/S bilayer calculated numerically for $\alpha_m
= 1/ \tau_m \Delta = 0.5$ (solid black line), $\alpha_m = 1$ (dashed red line), and
$\alpha_m = 3$ (dash-dotted blue line) for different values of the F layer thickness
$d_f$, $h = 4 \Delta$, $T = 0.1T_c$. Parameters of the F/S interface are $\gamma =
\gamma_{B2} = 0.01$.
(a): $d_f/\protect\xi_n = 0.5$, (b): $d_f/\protect\xi_n = 1$, (c): $%
d_f/\protect\xi_n = 2$, (d): $d_f/\protect\xi_n = 3$. For plots (c) and (d) the curves
with $\alpha_m = 3$ are not shown since they are of the order of unity at corresponding
scale. Inset of the plot (a): $N_f(E)$ dependence for $d_f/\protect\xi_n = 0.5$ for
higher values of $\alpha_m$; $\alpha_m = 5$ (solid black line), $\alpha_m = 7$ (dashed
red line), $\alpha_m = 15$ (dash-dotted blue line).} \label{DOS_a} \vspace{-5mm}
\end{figure}

Now we turn to the exact  numerical solution. The obtained energy dependencies of the DOS
at the free F boundary of the F/S bilayer are presented in
Figs.~\ref{DOS_a0},~\ref{DOS_h},~\ref{DOS_a}. The exchange field is chosen such that  $h
> \Delta$, which corresponds to the experimental situation.

Figure \ref{DOS_a0} shows  the DOS energy dependence for
different $d_f$ in the absence of magnetic scattering. At small
$d_f$ we observe the DOS double peak due to the Zeeman splitting
of the BCS peak at $E=\Delta$. Most probably in the experiments,
the BCS Zeeman splitted peak as presented in Fig.~\ref{DOS_a0} (a) will be seen
as a single peak due to many-body interaction effects, which
introduce a finite lifetime (damping) of the quasiparticles.
We also observe that at small $d_f$ and relatively
small exchange field $h$, full DOS turns to zero inside a minigap,
which vanishes with the increase of $d_f$.


The minigap also exists in the normal metal (N) DOS in the S/N bilayers. If the thickness
$d_n$ of the normal metal is larger than the coherence length, the characteristic scale
of the minigap is set by the Thouless energy, $E_{Th} = D_n/d_n^2$, where $D_n$ and is
the diffusion coefficient of the normal metal.\cite{GK1} In the F layer of the S/F
bilayer, the exchange field $h$ shifts the DOS for the two spin subbands in opposite
directions, therefore the critical value $h_c$ of the exchange field at which the minigap
in the spectrum closes can be roughly estimated as\cite{Ivanov}
\begin{equation}\label{Eth}
h_c \sim E_{Th}, \quad E_{Th} = D_f/d_f^2.
\end{equation}
This equation shows the qualitative tendency: for smaller $d_f$ a higher $h$ is needed to
close the minigap [see also Fig.~\ref{DOS_h} (a)]. The estimation, \Eq{Eth}, is only
valid in the absence of magnetic scattering, since $\tau_m$ also influences the
minigap\cite{Ivanov2} [see also Fig.~\ref{DOS_a} (a)].

\begin{figure}[tb]
\epsfxsize=8cm\epsffile{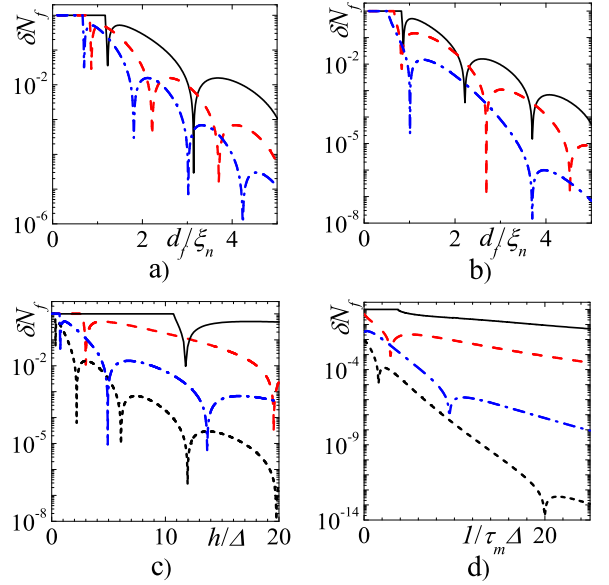} \vspace{-3mm} \caption{ (Color online) Dependence of
$\delta N_f$ as a function of the F layer thickness $d_f$ for different exchange fields
(a) and magnetic scattering times (b); dependence of $\delta N_f$ as a function of
exchange filed (c) and magnetic scattering time (d) for different $d_f$. The temperature
$T = 0.1 T_C$. Parameters of the F/S interface are $\gamma = \gamma_{B2} = 0.01$. (a): no
magnetic scattering ($1/\tau_m \Delta = 0$), $h/\Delta = 2$ (black solid line), $h/\Delta
= 4$ (red dashed line), $h/\Delta = 6$ (blue dash-dotted line); (b): $h/\Delta = 4$,
$\alpha_m = 1/\tau_m \Delta = 0$ (black solid line), $\alpha_m = 1$ (red dashed line),
$\alpha_m = 3$ (blue dash-dotted line); (c) no magnetic scattering, $d_f/\xi_n = 0.5$
(black solid line), $d_f/\xi_n = 1$ (red dashed line), $d_f/\xi_n = 2$ (blue dash-dotted
line), $d_f/\xi_n = 3$ (black short-dashed line); (d) $h/\Delta = 4$, $d_f/\xi_n = 0.5$
(black solid line), $d_f/\xi_n = 1$ (red dashed line), $d_f/\xi_n = 2$ (blue dash-dotted
line), $d_f/\xi_n  = 3$ (black short-dashed line).} \label{osc} \vspace{-5mm}
\end{figure}

In Fig.~\ref{DOS_a0} we also observe that after the minigap closes, the DOS at the Fermi
energy $N_f(0)$ rapidly increases to values larger than unity with further increase of
$d_f$; then it oscillates around unity while its absolute value exponentially approaches
unity [see also Fig.~\ref{osc}]. This is the well-known damped oscillatory behavior of the
DOS in F/S bilayers. Experimental evidence for such behavior was provided by Kontos {\it
et al}.\cite{Kontos_DOS} In the case of long enough ferromagnetic layer we also observe
the DOS peak at $E = h$, which was previously discussed in Ref.~\onlinecite{Buzdin_H}. A
similar effect was also discussed in N/F/S structures, where it was shown that a zero
energy peak appears in DOS if $E_{Th} = h$.\cite{resonant}

We also show in Fig.~\ref{DOS_a0} the  analytical approximation,
\Eq{Napprox}, which is in good agreement with the numerical result
for thick enough ferromagnetic layers. In the numerically obtained
curves the peak at $E = h$ is smeared because of finite $\gamma =
0.01$ for the transparent F/S interface at $x = d_f/2$.


In the absence of magnetic scattering  we rewrite the analytical
DOS expression, Eq.~(\ref{Napprox}), for $E \geq \Delta$ in the
following way,
\begin{align}
N_f(E) = 1 + \sum_\pm \frac{16 \Delta^2 \cos\left( \frac{2d_f}{\xi
_{f}}\sqrt{\frac{|E \pm h|}{h}}
\right)}{(E+\epsilon)(\sqrt{E+\epsilon} + \sqrt{2\epsilon})^2} e^{
-\frac{2d_f}{\xi _{f}}\sqrt{\frac{|E \pm h|}{h}}},\label{Exp}
\end{align}
where $\epsilon = \sqrt{E^2 - \Delta^2}$. We can clearly see the exponential asymptotic
of the peak at $E = h$ from the \Eq{Exp}. We should keep in mind that \Eq{Exp} is valid
for large $d_f/\xi_f$, but nevertheless we may qualitatively understand why we do not see
the peak at $E = h$ for small ratio of $d_f/\xi_f$: if this factor is small the variation
of the exponent $\exp\{-2(d_f/\xi_{f})\sqrt{|E - h|/h}\}$ near the point $E = h$ is also
small. The peak is observable only for $h$ of the order of a few $\Delta$. For larger
exchange fields the peak is very difficult to observe, since the energy dependent
prefactor of the exponent in \Eq{Exp} decays as $E^{-2}$ for $E \gg \Delta$.

Figure \ref{DOS_h} shows the DOS energy dependence for different
values of the exchange field $h$ in the absence of magnetic scattering.
For stronger exchange field the
minigap closes at smaller $d_f$, in qualitative correspondence
with \Eq{Eth}. From numerical calculations we obtain the following
condition [see also Fig.~\ref{osc} (c)], valid for $\tau_m^{-1} \sim 0$,\cite{minigap}
\begin{equation}\label{ETh1}
h_c \approx 0.77 E_{Th} \approx 2.71 \Delta \left(
\frac{\xi_n}{d_f} \right)^2.
\end{equation}

In Fig.~\ref{DOS_h} we also observe the peak at $E = h$; at large enough exchange fields
its amplitude can be much larger than the amplitude of the peak at $E = \Delta$ (see
Fig.~\ref{DOS_h} (d), blue dash-dotted curve). The existence of the DOS peak at $E = h$
gives a possibility to measure the exchange field directly in experiment by measuring the
F/S bilayer DOS in compounds with small magnetic scattering (since magnetic scattering is
smearing the peak, see below). For example, in Ref.~\onlinecite{small_h} were reported
exchange fields for Pd$_{1-x}$Ni$_x$ with different Ni concentration, obtained by a
fitting procedure. Considering Nb as a superconductor with $\Delta =$1.3 meV, we can
estimate the exchange field in Pd$_{1-x}$Ni$_x$: for 7\% of Ni fitting gives $h =$2.8
meV, which is 2.2$\Delta$, and for 11.5\% of Ni $h =$3.9 meV, which is
3$\Delta$.\cite{small_h} It is interesting to use direct measurements of the DOS peak at
$E = h$ to check these fitting predictions of Ref.~\onlinecite{small_h}.

Ferromagnetic metals
with exchange fields of the order of few $\Delta$ are crucially important for the fabrication of
SIFS junctions, valid for superconducting logic applications. Presently used ferromagnets have $h \gg \Delta$,
and therefore short oscillation length [see \Eq{x_osc}], which makes it difficult to control the F layer thickness.
In already existing SIFS structures the roughness is often larger than desired precision of $d_f$.\cite{Goldobin}
We hope that our results will trigger the experimental activity in finding ferromagnetic alloys with
$h$ of the order of few $\Delta$.

Figure \ref{DOS_a} shows the DOS energy dependence for different
values of magnetic scattering time. Similarly to Fig.~\ref{DOS_h},
for stronger magnetic scattering the minigap closes at smaller
$d_f$. Also the DOS peak at $E = h$, visible for long enough
ferromagnetic layer, is smeared. The analytical solution (not shown),
\Eq{Napprox}   also agrees quite well with the
numerical results for  $d_f \gg \xi_{f 1}$.

Although our results are obtained for weak ferromagnets, they can in certain cases be
extended for ferromagnets with strong exchange fields,  $h \gg \Delta$.  In the absence
of magnetic scattering the Usadel equation in energy representation, \Eq{Usadel}, can be
rewritten as,
\begin{equation}\label{Usadel_h}
\frac{i}{2} \frac{\partial ^{2}\theta _{f\uparrow (\downarrow
)}}{\partial y^{2}} = \left( \frac{E}{h} \mp 1\right) \sin
\theta_{f\uparrow (\downarrow )},
\end{equation}
where $y = x/\xi_f$ is the dimensionless coordinate. In the case of $h \gg \Delta$, we
can neglect the first term on the right hand side of \Eq{Usadel_h} to obtain the subgap
DOS. Thus, in that case the subgap structure scales with the length $\xi_f$ and for
example the results presented in Fig.~\ref{DOS_a0} for $h = 4\Delta$ also describe the
DOS in the case of a high exchange field if one scales $d_f$ correspondingly. This
procedure does not apply however for $h = 2 \Delta$ (see Fig.~\ref{DOS_h}), since in that  case  one  cannot
simply neglect the term $E/h$ in \Eq{Usadel_h}.

To show explicitly the aforementioned DOS oscillations at the
Fermi energy,\cite{Kontos_DOS} in Fig.~\ref{osc} we plot the
numerically calculated function
\begin{equation}  \label{r}
\delta N_f(d_f, h, \tau_m) = |1 - N_{f 0}|, \quad N_{f 0} = N_f(E
= 0).
\end{equation}

Using Eqs.~(\ref{DOS_bound})-\eqref{NapproxAB} and \eqref{r} we get the analytical
expression for the function $\delta N$, valid for $d_f \gg \xi_{f 1}$,
\begin{equation}  \label{ra}
\delta N = 32\biggl|\re\biggl[\frac{1}{(\sqrt{2 - \eta^2_0} +
1)^2} \exp\left( -p_0\frac{2 d_f}{\xi_f}\right) \biggr]\biggr|,
\end{equation}
where
\begin{subequations}
\begin{align}
p_0 &= \sqrt{2/h}\sqrt{ih + 1/\tau_m},
\\
\eta_0 &= (1/\tau_m) (ih + 1/\tau_m)^{-1}.
\end{align}
\end{subequations}
At vanishing magnetic scattering we obtain,
\begin{equation}  \label{r0}
\delta N = \frac{32}{3 + 2\sqrt{2}} \left| \cos\left(\frac{2
d_f}{\xi_{f}}\right) \exp\left(-\frac{2 d_f}{\xi_{f}}\right)
\right|,
\end{equation}
in which case the characteristic lengths of decay and oscillations
are equal to $\xi_f$.

The dependence of $\delta N_f(d_f)$ on the ferromagnetic layer thickness $d_f$ at
different values of exchange field and magnetic scattering time is presented in
Fig.~\ref{osc} (a) and (b). From Fig.~\ref{osc} (a) we can see that with increasing
exchange field $h$ the minigap closes at smaller $d_f$ in agreement with \Eq{ETh1}, the
period of the DOS oscillations at the Fermi energy decreases, and the damped exponential
decay occurs faster. This is easy to see from \Eq{r0}, since in the absence of magnetic
scattering $\delta N$ depends on $h$ only as a function of $\xi_f$.

From Fig.~\ref{osc} (b) we can see that with increasing $\alpha_m = 1/\tau_m \Delta$ the
period of the DOS oscillations on the contrary increases, although the minigap also
closes at smaller $d_f$ and the damped exponential decay occurs faster. To understand
this behavior we rewrite here the decay and oscillation lengths, \Eq{xif12}, at the Fermi
energy,
\begin{subequations}\label{xi0f12}
\begin{align}
\frac{1}{\xi_{f1}} &= \frac{1}{\sqrt{D_f}} \sqrt{\sqrt{h^2 +
\frac{1}{\tau_m^2}} + \frac{1}{\tau_m}},
\\
\frac{1}{\xi_{f2}} &= \frac{1}{\sqrt{D_f}} \sqrt{\sqrt{h^2 +
\frac{1}{\tau_m^2}} - \frac{1}{\tau_m}}.
\end{align}
\end{subequations}
We see from these equations that with increasing $\alpha_m$ the length of decay $\xi_{f
1}$ decreases, while the length of oscillations $\xi_{f 2}$ increases.

The dependencies of $\delta N_f$ on exchange field and magnetic scattering time are
presented in Fig.~\ref{osc} (c) and (d), correspondingly. In Fig.~\ref{osc} (c) we see
oscillations of the DOS at the Fermi energy $\delta N_f(h)$ around unity with increasing
exchange field in the absence of magnetic scattering. In Fig.~\ref{osc} (d) we show the
function $\delta N(\tau_m)$. It is interesting to note that its behavior can be
both oscillatory and also monotonous. When the parameter $\alpha_m$ increases starting
from the minigap state (black solid curve) it is totally monotonous: increasing
$\alpha_m$ the minigap closes and the DOS starts to increase to unity, but never
overshoots unity [we checked this up to $\alpha_m = 80$, see also Fig.~\ref{DOS_a} (a)].
If we start from the state where the minigap is already closed, we first observe
oscillations, but then again a switch to monotonous behavior. For intermediate F layer
thicknesses we see just one oscillation and then DOS monotonously approaches unity (we
checked this up to $\alpha_m = 80$), while for thicker ferromagnets ($d_f/\xi_n = 3$) we
observe two oscillations and then a monotonic behavior.

The dependencies $\delta N(h)$ and $\delta N(\tau_m)$ can be
important if in the experiment the material properties of the
ferromagnetic interlayer, {\it i.e.} exchange field $h$ and
magnetic scattering time $\tau_m$ can vary with some external
parameter, for example temperature, magnetic field, etc.

Before turning to the calculation of the  CVC we discuss briefly a recent experiment
\cite{Beasley} in which a pronounced double peak in the DOS of  Ni/Nb bilayers were
reported. This double peak cannot be explained within our model  based on the  Zeeman
splitting. The reason for the double peak in Ref.~\onlinecite{Beasley} remains
controversial. In Ref.~\onlinecite{CottetLinder} it was numerically fitted by adding an
extra parameter to the model, characterizing spin-active interfaces. However, this fit is
far from being satisfactory. Nevertheless there is  another feature of the DOS observed
in Ref.~\onlinecite{Beasley} which can be explained within our model:  by increasing
$d_f$ the ``normal'' peak at $E = \Delta$  [which is the BCS Zeeman split peak in
Fig.~\ref{DOS_a0}(a)] is  ``inverted''  [Fig.~\ref{DOS_a0}(b) and (c)] and becomes
``normal'' again [Fig.~\ref{DOS_a0}(d)] as $d_f$ is further increased. According to our
model,  at $E = \Delta$ \Eq{Exp} reduces to the following expression,
\begin{equation}
N_f(\Delta) = 1 + 16 \sum_\pm \cos \biggl ( \frac{2 d_f}{\xi_f}
\sqrt{\frac{h \pm \Delta}{h}} \biggr ) e^{-\frac{2d_f}{\xi
_{f}}\sqrt{\frac{h \pm \Delta}{h}}}.\label{ExpD}
\end{equation}
This expression  explains the inversion of the peak at $E =
\Delta$ as a function  of  $d_f$.  The peak is ``normal''
(``inverted'') if the DOS at $\Delta$ is larger (smaller) than
unity.  This variation is due to the sign of the cosine function
in \Eq{ExpD}, which depends on the $d_f/\xi_f$ ratio.


\section{Current-voltage characteristics of a SIFS junction}\label{CVC_SIFS}

In this section we calculate the current-voltage characteristics (CVC) of a SIFS junction
at low temperature, $T = 0.1 T_c$, using \Eq{I(V)} and DOS $N_f(E, d_f, h, \tau_m)$
numerically obtained in the previous section.

\begin{figure}[t]
\epsfxsize=7.5cm\epsffile{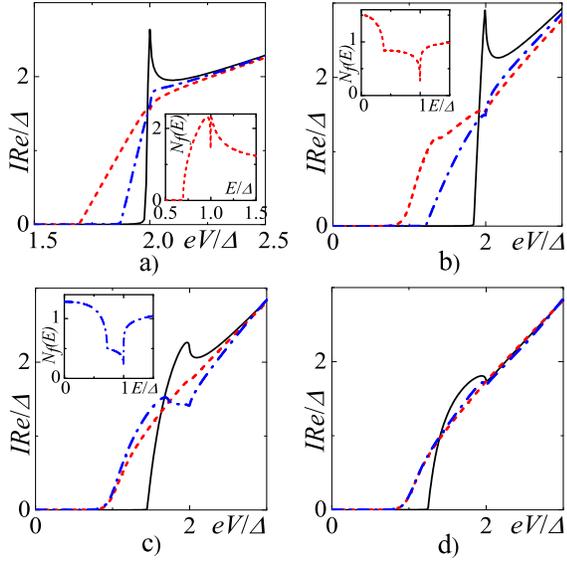} \vspace{-3mm}
\caption{(Color online) Current-voltage characteristics of a SIFS
junction in the absence of magnetic scattering for different
values of the F-layer thickness $d_f$. The temperature $T = 0.1
T_c$. The exchange field $h = 0$ (black line, which corresponds to
the case of a SINS junction), $h/\Delta = 2$ (blue dash-dotted
line), and $h/\Delta = 4$ (red short-dashed line). (a):
$d_f/\xi_n=0.5$; (b): $d_f/\xi_n=1$, (c): $d_f/\xi_n=2$, and (d):
$d_f/\xi_n=3$. Insets in (a), (b) and (c) are explained in the text.}
\label{IVa0}\vspace{-5mm}
\end{figure}

Figure \ref{IVa0} shows the CVC  of a SIFS junction in the absence
of magnetic scattering. For comparison we also present  the CVC of
a SINS tunnel junction, {\it i.e.} a junction with a normal metal
interlayer instead of a ferromagnet ($h = 0$). SINS structures
were studied previously  in Ref.~\onlinecite{GK1}. We observe
several features of SIFS CVC  which are the signatures of the
proximity effect in the S/F bilayer.

For thin enough F layer
we observe the ``kink'' on the CVC at $eV \approx
2\Delta$ [Fig.~\ref{IVa0} (a)], which corresponds to the
case when the DOS $N_f(E)$ exhibits a pronounced minigap. The corresponding
DOS energy dependence ($h/\Delta = 4$, $d_f/\xi_n = 0.5$) is shown
in the inset. We can also see that for a
certain range of parameters the CVC of a SIFS junction exhibit a
non-monotonic ``wave'' behavior. We can observe it for $h/\Delta =
4$ (red short-dashed line) in Fig.~\ref{IVa0} (b) and for
$h/\Delta = 2$ (blue dash-dotted line) in Fig.~\ref{IVa0} (c).
This behavior corresponds to the case when the DOS $N_f(E)$
minigap is already closed and the $N_f(0)$ at the Fermi energy is
larger than unity. The corresponding DOS energy dependencies are
presented in the insets of the plots \ref{IVa0} (b) and (c).

\begin{figure}[t]
\epsfxsize=7.5cm\epsffile{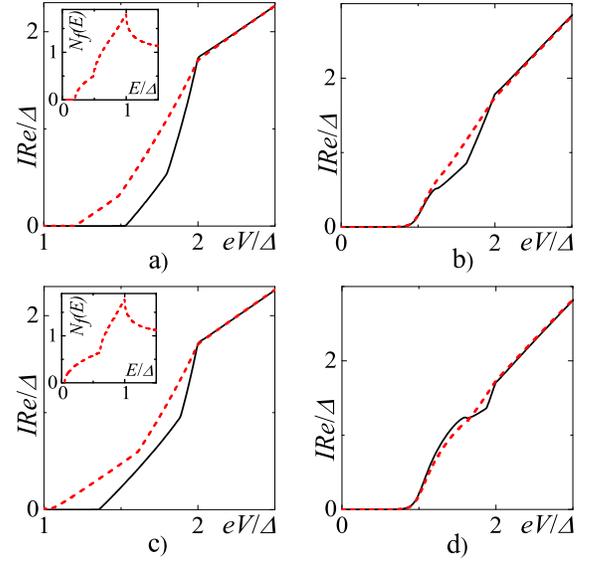}\vspace{-3mm}
\caption{(Color online) Current-voltage characteristics of a SIFS
junction for $1/\tau_m \Delta = 1$ (black solid lines) and
$1/\tau_m \Delta = 3$ (red short-dashed lines) for different
values of the F-layer thickness $d_f$. The temperature $T
= 0.1 T_c$. The exchange field $h/\Delta = 2$: plots (a) and (b)
and $h/\Delta = 4$: plots (c) and (d). The thickness
$d_f/\xi_n=0.5$: plots (a) and (c) and $d_f/\xi_n=1$: plots (b)
and (d). Insets in (a) and (c) are explained in the text.}
\label{IVa1}\vspace{-5mm}
\end{figure}

At large enough $d_f$ and exchange fields the DOS $N_f(E) \approx
1$ and the current, Eq.~(\ref{I(V)}), is given by the same
equation as the current in the NIS tunnel junction,
\begin{equation}\label{I(V)long}
I = \frac{1}{e R} \int_{-\infty}^{\infty}dE N_s(E) \left[ f(E -
eV) - f(E) \right].
\end{equation}
At small temperature $T \ll T_c$, this equation is well
approximated by taking $T = 0$,
\begin{align}
I &= \Theta(eV - \Delta) \frac{1}{e R} \int_{\Delta}^{eV}\frac{E
\; dE}{\sqrt{E^2 - \Delta^2}}\nonumber
\\
&= \Theta(eV - \Delta) \frac{\sqrt{(eV)^2 -
\Delta^2}}{eR},\label{I(V)T0long}
\end{align}
The red short-dashed line in Fig.~\ref{IVa0} (d) ($d_f/\xi_n = 3$, $h/\Delta = 4$, no
magnetic scattering) almost coincides with this result except for the small region $eV
\approx \Delta$, since for our numerically calculated curves we fix temperature $T = 0.1
T_c \ll T_c$.

Figure \ref{IVa1} shows current-voltage characteristics of a SIFS junction in case of
finite magnetic scattering. Here for thin F layers we observe a ``double-kink''
structure, see Fig.~\ref{IVa1} (a) and (c). It corresponds to the DOS $N_f(E)$ with small
minigap and finite subgap value smaller than unity. Such a DOS structure is typical in
the presence of magnetic scattering and thin enough ferromagnetic interlayer, see
Fig.~\ref{DOS_a} (a). The corresponding DOS energy dependencies are presented in the
insets of the plots \ref{IVa0} (a) [$1/\tau_m \Delta = 3, h/\Delta = 2, d_f/\xi_n = 0.5$]
and (b) [$1/\tau_m \Delta = 3, h/\Delta = 4, d_f/\xi_n = 0.5$]. For finite magnetic
scattering the non-monotonic features of CVC are smeared. We do not show the curves for
$d_f/\xi_n \geq 2$, since they do not significantly differ from the curves obtained from
Eq.~(\ref{I(V)T0long}).

Figures \ref{IVa0} (a) and \ref{IVa1} (a) show that for thin enough ferromagnetic layer
the current has an onset in the interval
$\left[ \Delta, \; 2 \Delta \right]$ (for temperatures $T \ll T_C$).
The value of this onset, according to \Eq{I(V)T0}, is $\Delta + E_g$,
where $E_g$ is the DOS minigap, $0 < E_g < \Delta$. Increasing
exchange field, magnetic scattering and/or F layer thickness, the minigap closes and the current
turns to zero at $eV < \Delta$, having an onset at $eV = \Delta$.
The dependence of the minigap $E_g$ on the parameters characterizing the material
properties of the ferromagnetic interlayer is discussed in Section~\ref{DOS_FS}.

We conclude that we observe interesting features in the SIFS CVC if the DOS $N_f(E)$ near
the insulating barrier has a nontrivial shape in the subgap region.
In case when $N_f \approx 1$, these features disappear and the CVC coincide with those of
the NIS tunnel junction, \Eq{I(V)long}.


\section{Macroscopic Quantum Tunneling in a SIFS junction}\label{sec-MQT}

\begin{figure}[t]
\epsfxsize=8.5cm\epsffile{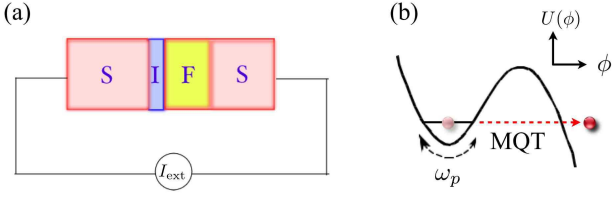} \vspace{-3mm} \caption{(Color
online) (a) Schematic of a current biased SIFS Josephson junction.
$I_\mathrm{ext}$ is the external bias current. (b) Potential
$U(\phi)$ v.s. the phase difference  $\phi$ between two
superconductors. $\omega_p$ is the Josephson plasma frequency of
the junction. } \label{MQT} \vspace{-5mm}
\end{figure}

In this section, motivated by experimental studies on the (thermal and quantum)
switching~\cite{rf:Aprili} and quantum coherent oscillations~\cite{rf:Feofanov} in SFS
and SIFS junctions, we calculate the MQT escape rate in a current biased SIFS junction as
shown in Fig.~\ref{MQT} (a). The CVC obtained in the previous section enable us to
investigate the influence of the quasiparticle dissipation on MQT.

It is important to note that MQT can be used as a measurement process of a
superconducting phase qubit.~\cite{phase-qubit}
Thus, the calculation of the MQT rate by taking into account the
quasiparticle dissipation will be very important for analyzing the
fidelity of measurement process for phase qubits.
In the following calculation, for simplicity, we have ignored the
influence from an environmental circuit on MQT which can be experimentally
reduced by a noise filtering technique.\cite{noise_filtering}

The partition function of a junction can be described by an
imaginary-time functional integral over the macroscopic variable
(the phase difference $\phi$ between two
superconductors),~\cite{rf:Zaikin,rf:Ambegaokar,rf:Kawabata1,rf:Kawabata2} {\it
i.e.},
\begin{equation}\label{eqn:MQT1}
Z = \int D \phi (\tau) \exp \left(
  - \frac{S_{\mathrm{eff}}[\phi]}{\hbar}
\right).
\end{equation}
%

In the strong insulating barrier limit, {\it i.e.}, $\gamma_{B1}
\gg 1$, the effective action $S_{\mathrm{eff}}$ is given by
\begin{subequations}
\begin{align}
S_{\mathrm{eff}}[\phi] &= \int_{0}^{\hbar \beta} d \tau \left[
   \frac{M}{2}
   \left(
   \frac{\partial \phi ( \tau) }{\partial \tau}
   \right)^2
   +
   U(\phi)
\right] + S_\alpha[\phi],\label{eqn:MQT2}
\\
S_\alpha[\phi] &= - \int_{0}^{\hbar \beta} d \tau \int_{0}^{\hbar
\beta} d \tau'
  \alpha (\tau - \tau') \cos \frac{\phi(\tau) - \phi (\tau') }{2}.
  \label{eqn:MQT3}
\end{align}
\end{subequations}
In this equation, $\beta = 1 /k_B T$, $M = C \left( \hbar/2
e\right)^2 $ is the mass ($C$ is the capacitance of the junction)
and the potential $U(\phi)$ can be described by a tilted washboard
potential [Fig.~\ref{MQT} (b)], {\it i.e.},
\begin{equation}
 U(\phi)
 =
-E_J \left[
   \sgn(I_c) \cos \phi +  y \phi
\right],\label{eqn:MQT4}
\end{equation}
with  $y \equiv I_{ext} / | I_c |$, where $E_J=\hbar |I_c |/ 2e$
is the Josephson coupling energy, $I_c$ is the Josephson critical
current, and $ I_{ext}$ is the external bias current,
respectively. The dissipation kernel $\alpha(\tau)$ is related to
the quasiparticle current $I$ under constant bias voltage $V$ by
\begin{equation}
\alpha(\tau) = \frac{\hbar }{e} \int_0^\infty \frac{d \omega}{2
\pi}  \exp \left(  -\omega \tau \right) I \left( V=\frac{\hbar
\omega}{e} \right),\label{eqn:MQT5}
\end{equation}
at zero temperature.~\cite{rf:Zaikin,rf:Ambegaokar,rf:Kawabata1}

As clearly seen from Figs. 6 and 7, the CVC has a gap structure due to the isotropic
superconducting gap in left superconductor electrode. In such a case, the dissipation
kernel $\alpha$ decays exponentially as a function of imaginary time $\tau$ for $ |\tau|
\gg \hbar / \Delta$. The typical dynamical time scale of the macroscopic variable $\phi$
is of the order of the inverse Josephson plasma frequency $\omega_p= \sqrt{2e |I_c| /
\hbar C}( 1-y^2)^{1/4}$ which is much smaller than $\Delta$. Thus, the phase varies
slowly with the time scale given by $\hbar /\Delta$, and we can expand $\phi(\tau) - \phi
(\tau')$ in Eq. (\ref{eqn:MQT3}) about $\tau=\tau'$. This gives
\begin{equation}
S_\alpha[\phi] \approx \frac{\delta C}{2} \int_{0}^{\hbar \beta} d
\tau
   \left[
   \frac{\hbar}{2e}
   \frac{   \partial \phi ( \tau) }{\partial \tau}
   \right]^2,\label{eqn:MQT6}
\end{equation}
where
\begin{equation}
\delta C =
 2
\left(  \frac{2 e}{\hbar }  \right)^2 \int_0^\infty d\tau
\alpha(\tau) \tau^2.\label{eqn:MQT7}
\end{equation}
Hence, the dissipation action $S_\alpha$ acts as a kinetic term so
that the effect of the quasiparticles results in an increase of
the capacitance, $C \to C + \delta C \equiv C_{ren}$.

In case of a thin ferromagnetic layer ($d_f = 0.5\xi_n$) we numerically obtain $\delta C
\approx \hbar/\Delta R$ for CVC presented in Figs.~\ref{IVa0} (a), ~\ref{IVa1} (a), (c)
[``kink'' and ``double-kink'' structures]. For thick ferromagnetic layer we can use
\Eq{I(V)T0long} to calculate $\delta C$,
\begin{equation}
\delta C = \frac{4 \hbar}{\pi \Delta R}\int_0^{\infty} dx \; x^2
\int_1^{\infty} dz \; e^{-xz} \sqrt{z^2 - 1} \approx \frac{2
\hbar}{\Delta R}.
\label{eqn:capacitance}
\end{equation}
For intermediate $d_f$ we numerically find $\delta C \approx (1 \div 2) \; \hbar/\Delta
R$. Considering Nb as a superconductor ($\Delta =1.3$ meV) we therefore obtain $\delta C
\approx (0.5 \div 1) r^{-1}$ pF, where $r$ is the junction resistance $R$ in $\Omega$. To
insure a small dissipative correction of capacitance, $\delta C \ll C$ we have a
constraint,
\begin{equation}\label{RC}
RC \gg \hbar/\Delta,
\end{equation}
{\it i.e.} the typical time constant $RC$ of a SIFS junction should be much larger than
the dynamical damping scale for the dissipation kernel $\alpha(\tau)$. For example, in
Ref.~\onlinecite{Weides} the following parameters of a
Nb/Al$_2$O$_3$/Ni$_{0.6}$Cu$_{0.4}$/Nb SIFS junction were reported, $C =$800 pF and $R
=$55 m$\Omega$, which correspond to $\delta C \approx$ 10$\div$18 pF. Thus, even for the
low resistive tunnel barrier in Ref.~\onlinecite{Weides}, we have the condition \Eq{RC}
satisfied.

In order to see the effect of the quasiparticle dissipation on macroscopic quantum
dynamics, we will investigate MQT in current-biased SIFS junctions. The MQT escape rate
$\Gamma$  from the metastable potential at zero temperature is given by~\cite{rf:MQT1}
\begin{equation}
\Gamma = \lim_{\beta \to \infty} \frac{2}{\beta} \mbox{ Im}\ln Z.
  \label{eqn:MQT8}
\end{equation}
By using the Caldeira and Leggett theory,~\cite{rf:Caldeira} the
MQT rate is approximated as
\begin{equation}
\Gamma = \frac{\hat{\omega}_p}{2 \pi}\sqrt{120 \pi B }\  \exp (
-B),\label{eqn:MQT9}
\end{equation}
where
\begin{equation}
\hat{\omega}_p=  \sqrt{\frac{2 e  |I_c| }{\hbar
C_\mathrm{ren}}}(1-y^2)^{1/4},
  \label{eqn:MQT10}
\end{equation}
is the renormalized Josephson plasma frequency and $B =
S_{\mathrm{eff}}[\phi_B]/\hbar$ is the bounce exponent, that is
the value of the action $S_{\mathrm{eff}}$ evaluated along the
bounce trajectory $\phi_B(\tau)$. The analytic expression for the
bounce exponent is given by
\begin{equation}
B=\frac{12}{5 e}
  \sqrt{ \frac{\hbar}{2e } |I_c| C_\mathrm{ren}}
  \left(
  1 -
 y^2
\right)^{5/4} .
  \label{eqn:MQT11}
\end{equation}
%

\begin{figure}[t]
\epsfxsize=8.5cm\epsffile{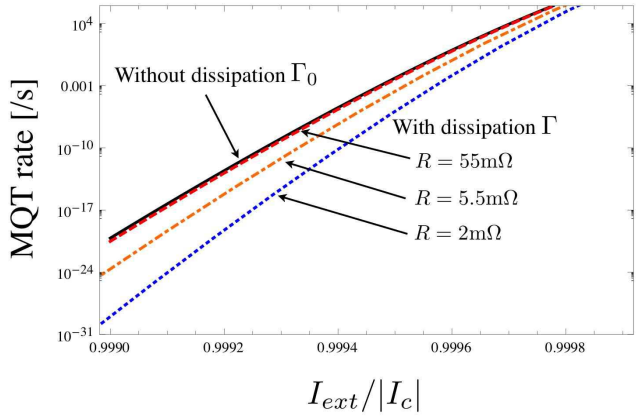} \vspace{-3mm} \caption{(Color online) The MQT escape
rate for a current-biased SIFS junction as a function of the bias current $I_{ext}$ for several values of the junction resistance $R$.
$\Gamma_0$ (black solid line) and $\Gamma$ (red dashed, yellow dot-dashed, and blue dotted lines) are  the MQT escape rate
without and with quasiparticle dissipation, respectively. Parameters are $C=800$ pF, $\Delta=1.3$ meV, and $|I_c|=500$ $\mu$A.\cite{Weides} } \label{MQTrate}
\vspace{-5mm}
\end{figure}

At high temperatures, the thermally activated decay dominates the escape process.
Then the escape rate is given by the Kramers formula,
$\Gamma=(\hat{\omega}_p/ 2 \pi ) \exp ( - U_0 / k_B T)$,
where $U_0$ is the barrier height.~\cite{rf:MQT1}
Below the crossover temperature $T^*$, the escape process is dominated by MQT.
In the low dissipative (underdamping) cases, $T^* $ is
approximately given by~\cite{rf:MQT1,rf:Kato}
\begin{equation}
T^* =\frac{ \hbar \hat\omega_p(y= \langle y \rangle) }{ 2 \pi k_B},
\end{equation}
Here $\langle y \rangle= \int_0^1 d y P(y) y$ is the average switching current, where $P(y)$ is the switching current distribution which is related to the escape rate $\Gamma$ as\cite{noise_filtering}
\begin{eqnarray}
P(y)=\frac{1}{v}
 \Gamma (y) \exp
\left[
- \frac{1}{v}
\int_0^{y} \Gamma (y') d y'
\right]
.
\end{eqnarray}
In this equation, $v \equiv \left| d y / d t \right| $ is the sweep rate of the external bias current.
Importantly, $T^*$ is reduced in the presence of dissipative effects.~\cite{rf:Caldeira}

By using Eqs. (\ref{eqn:capacitance}) and (\ref{eqn:MQT9}), we
calculate $\Gamma$ and compare it with the case without the quasiparticle dissipation.
In Fig. 9, we numerically plot $\Gamma$ and $\Gamma_0$ for $C=800$ pF and
$|I_c|=500$ $\mu$A\cite{Weides} and several values of $R$, where $\Gamma_0$ is the MQT escape rate in the
absence of the quasiparticle dissipation [$C_\mathrm{ren} \to C$ in \Eq{eqn:MQT9}].
As seen in this figure, $\Gamma$  shows strong dependence on the junction
resistance $R$, and $\Gamma$  is almost identical to $\Gamma_0$ in the
case of large $R$, e.g., $R=55$ m$\Omega$ ($\gg \hbar / C \Delta  \approx 0.63$ m$\Omega$)
which corresponds to the actual SIFS junction.\cite{Weides}
We also calculate $T^*$ for a realistic case ($R=55$ m$\Omega$) and found
that $T^*=7.4 $mK for the dissipative case ($C_\mathrm{ren}=C+\delta C$)
and $T^*=7.5 $mK for the dissipation-less case ($C_\mathrm{ren}=C$).
As expected, the $T^*$ suppression is small enough to allow experimental
observations of MQT.
Thus we can conclude that the influence of the quasiparticle dissipation
on the macroscopic quantum dynamics of
 SIFS junction is very small for the case when the condition \Eq{RC} is hold.
This fact strongly suggests the great advantage of realistic SIFS
junctions for qubit applications. The smallness of the quasiparticle dissipation
in SIFS junctions is due to the superconducting gap in the left S electrode and the
strong insulating barrier ($\gamma_{B1} \gg 1$) between the left S and F layers.

It is important to note that such a weak quasiparticle-dissipation nature of
MQT has been also predicted in $\pi$-junctions based on a S/ferromagnetic insulator (FI)/S
junctions.\cite{rf:Kawabata2, rf:Kawabata3}
However no ferromagnetic insulator based Josephson junctions have been experimentally realized at present.
On the other hands, the fabrication of SIFS  junction is easily
realized based on the current fabrication technology.\cite{Weides, Weides2, Weides3}


\section{Conclusion}\label{Concl}

We have developed a quantitative theory, which describes the properties of the DOS and
the current-voltage characteristics of a SIFS junction in the dirty limit. We considered
the case of a strong insulating barrier in a SIFS junction such that the left S layer and
the right F/S bilayer are decoupled. In this case we can obtain the current-voltage
characteristics of a SIFS junction in the framework of standard tunnel theory. In order
to calculate quasiparticle current we first calculated the DOS in the ferromagnetic layer
of the F/S bilayer. We described the DOS behavior as a function of parameters
characterizing properties of the ferromagnetic layer. In our theory we consider three
such parameters: thickness of the ferromagnetic layer $d_f$, exchange filed $h$, and
magnetic scattering $\tau_m$. We have discussed the DOS properties paying special
attention to the DOS oscillations at the Fermi energy. We have also proposed to measure
the exchange field in experiments on weak ferromagnets by measuring the DOS peak at $E = h$.
We compared the results, obtained with a self-consistent numerical method, with a known
analytical DOS approximation, which is valid when the ferromagnetic layer is thick
enough.

Using the numerically obtained DOS we have calculated the current-voltage characteristics of a SIFS junction
and have observed features which are the signatures of the proximity effect in the S/F
bilayer. We showed that there exists typical shape patterns of current-voltage
characteristics related to the typical DOS structures in the ferromagnetic interlayer.

Finally, we have calculated the macroscopic quantum tunneling
escape rate for the current biased SIFS junctions by taking into
account the dissipative correction due to the quasiparticle
tunneling. Based on this we concluded that the influence of the
quasiparticle dissipation on the macroscopic quantum dynamics of
SIFS junctions is small, which is a great advantage of SIFS
junctions for qubit applications compared to other types of ferromagnetic $\pi$ - junctions.


\begin{acknowledgments}
The authors thank A.~I.~Buzdin, E.~Goldobin, D.~A.~Ivanov,
A.~V.~Vedyayev, and M.~Weides for
useful discussions. This work was supported by NanoSCIERA project
Nanofridge, ANR DYCOSMA, RFBR project N11-02-12065, JST-CREST, a
Grant-in-Aid for Scientific Research from the Ministry of
Education, Science, Sports and Culture of Japan (Grant No.
22710096), the Invitation Program for Foreign Young Researchers in
the G-COE program ``Education and research center for material
innovation'', and the Spanish MICINN (Contract No. FIS2008-04209).
F.S.B. thanks Intramural Special Project (Ref. 2009601036).
A.A.G. thanks Dutch FOM for support.
A.S.V. acknowledge the hospitality of Nanomaterials theory group, AIST,
during his stay in Japan.
\end{acknowledgments}

\end{document}